\documentclass[aps,prd,twocolumn,amsmath,amssymb,amsfonts,nofootinbib,superscriptaddress,altaffilletter]{revtex4-2}

\usepackage{graphicx}
\usepackage{hyperref}
\hypersetup{
  bookmarksopen=true
}
\usepackage{ulem}
\usepackage{amssymb}
\usepackage{amsmath}
\usepackage{color}
\usepackage{supertabular}
\usepackage{dcolumn}
\usepackage[mathscr]{eucal}
\usepackage{mathrsfs}
\usepackage{siunitx}

\def\sci#1#2{#1\times10^{#2}}

\begin{document}

\title{Early release of the expanded atlas of the sky in continuous gravitational waves}

\author{Vladimir Dergachev}
\email{vladimir.dergachev@aei.mpg.de}
\affiliation{Max Planck Institute for Gravitational Physics (Albert Einstein Institute), Callinstrasse 38, 30167 Hannover, Germany}
\affiliation{Leibniz Universit\"at Hannover, D-30167 Hannover, Germany}

\author{Maria Alessandra Papa}
\email{maria.alessandra.papa@aei.mpg.de}
\affiliation{Max Planck Institute for Gravitational Physics (Albert Einstein Institute), Callinstrasse 38, 30167 Hannover, Germany}
\affiliation{Leibniz Universit\"at Hannover, D-30167 Hannover, Germany}

\begin{abstract}
We present the early release of the atlas of continuous gravitational waves covering frequencies from 20\,Hz to 1500\,Hz and spindowns from $\sci{-5}{-10}$ to $\sci{5}{-10}$\,Hz/s. Compared to the previous atlas release we have greatly expanded the parameter space, and we now also provide polarization-specific data - both for signal-to-noise ratios and for the upper limits. Continuous wave searches are computationally difficult and take a long time to complete. The atlas enables new searches to be performed using modest computing power. To allow new searches to start sooner, we are releasing this data early, before our followup stages have completed.
\end{abstract}

\maketitle

\section{Introduction}

\begin{figure}[!htbp]
\includegraphics[width=3.3in]{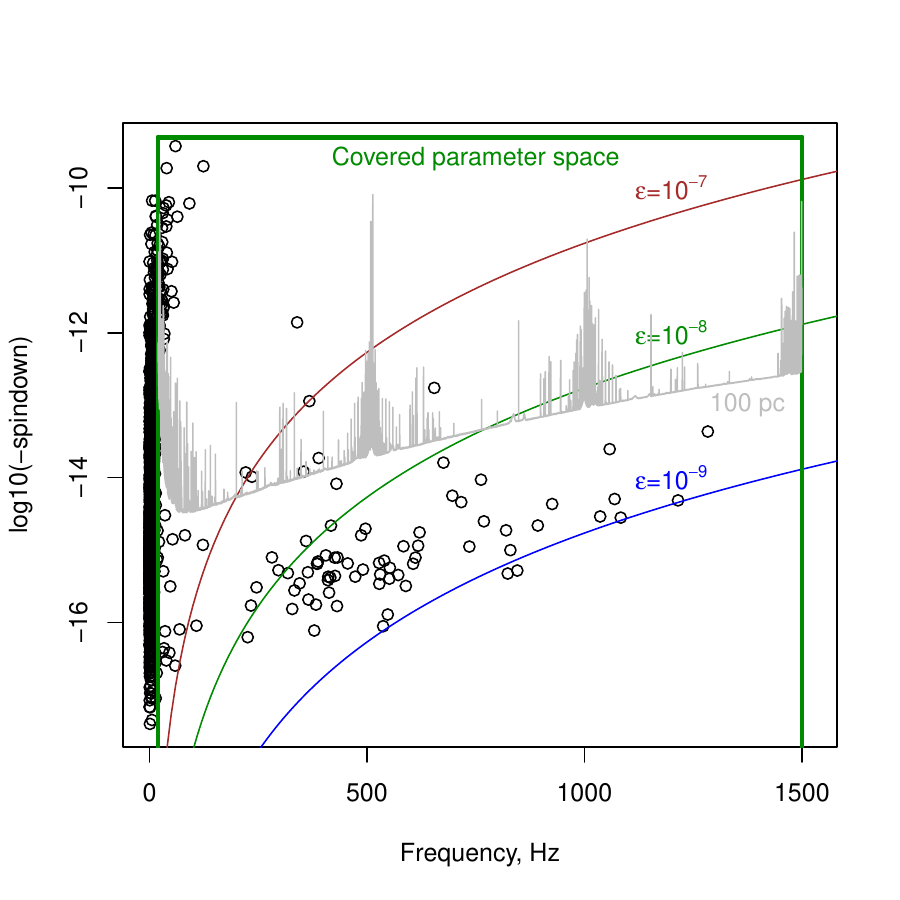}
\caption[Pulsar parameter space]{
\label{fig:pulsars}
This plot shows the expected gravitational wave frequency and frequency-derivative of isolated neutron stars from the ATNF catalog \cite{ATNF} (circles). The curves show the frequency/frequency-derivative combinations for a gravitar (a neutron star loosing energy solely due to gravitational wave emission) for different values of the ellipticity. Following \cite{ellipticity} we observe that known pulsars have spindown values that lie at or above the curve $\epsilon=10^{-9}$. For gravitars this would imply that the minimum ellipticity is around $10^{-9}$. 
The region above the gray trace and enclosed by the box of green straight lines shows the frequency/frequency-derivative combinations of optimally oriented gravitars at 100\,pc, that our search could detect. The frequency-derivative of a detectable gravitar is proportional to the square of the distance, so if the distance increases by a factor of 10, correspondingly the grey trace moves up by a factor of 100.
}
\end{figure}

This paper presents the early release of the expanded atlas of continuous gravitational waves. The atlas contains  upper limits and signal-to-noise ratios (SNR) as a function of signal frequency and polarization, and source sky position. Compared to the first atlas \cite{o3a_atlas1}, we have greatly expanded the searched parameter space (Fig. \ref{fig:pulsars}) and provide polarization-specific information, while maintaining comparable sensitivity.

The atlas is constructed from the results of the first two stages of the Falcon search \cite{o3a_atlas1, O2_falcon, O2_falcon2, O2_falcon3}. Subsequent stages do not improve the sensitivity of the analysis, but are used to identify potential signals. While we wait for these followups to complete, we are releasing the atlas data so that other scientists can use it earlier.

The data in the atlas can be used to find signals from neutron stars with equatorial deformations of $~\approx 100$\,\si{\um} up to 200\,pc away. Such sources could well exist in our immediate galactic neighbourhood. One can also search and place bounds on signals from other sources, such as unstable r-modes and boson condensates \cite{bo_rmodes, superradiance}.

The atlas contains seven billion records. It can be analyzed on a small personal computer thanks to the MVL library \cite{RMVL, MVL}. The atlas data, example code, and documentation can be found online \cite{data}.  We also recommend to consult \cite{o3a_atlas1} and \cite{functional_upper_limits}.

In the following sections, we present a brief overview of the key scientific aspects of the atlas.

\section{Continuous wave signals}

Falcon performs loosely coherent searches for bundles of signals with frequency evolution $f(t)$ of the IT2 type \cite{O2_falcon2} plus a small higher order perturbation:
\begin{equation}
f(t)=f_0+(t-t_0)f_1+(t-t_0)^2f_2/2+\sum_{n=3}^\infty (t-t_0)^n\frac{f_n}{n!},
\label{eq:freqEvolution}
\end{equation}
This search covers signals with $f_0$ from 20 to 1500\,Hz, $\left|f_1\right|\le \sci{5}{-10}$\,Hz/s, $\left|f_2\right|\le \sci{4}{-20}$\,Hz/s$^2$ and higher order terms:
\begin{equation}
\left|f_{n \ge 3}\right| \le \frac{\sci{2.9}{-6}\,\textrm{Hz}\cdot n!}{\left(\sci{1.5}{7}\,\textrm{s}\right)^n}.
\end{equation}
The reference time (GPS epoch) is $t_0=1246070000$ (2019 Jul 2 02:33:02 UTC).

It is expected that a rotating neutron star with an equatorial quadrupolar deformation will emit continuous gravitational waves. The amplitude of these waves grows as the first power of the deformation parameter $\epsilon$ (``ellipticity'') and with the second power of the frequency. 

Theory predicts that the neutron star crust can support quadrupolar deformations of $10^{-6}$ and perhaps even larger for neutron stars with exotic composition  \cite{Johnson-McDaniel:2012wbj,Gittins:2020cvx,Gittins:2021zpv}. However, signals like this have not yet been detected \cite{keith_review, lvc_O3_allsky, lvc_O3_allsky2, EatHO3a, LIGOScientific:2021hvc, Pagliaro:2023bvi, aashish_keith_loosely_coherent}, suggesting that extant neutron stars have much smaller ellipticities.

The spindown range for this search was chosen to cover rotating neutron stars with ellipticities of $10^{-7}$ up to $1500$\,Hz plus an extra margin to allow for increased spindown due to magnetic fields or kinematic effects. The highest frequency searched -- 1500\,Hz -- is above the gravitational wave frequency expected from the fastest known pulsar.  The detectors' sensitivity drops sharply near 20\,Hz, which we chose as the lowest search frequency. 

The chosen parameter space also covers a range of boson-condensate signals, with the frequency depending on the particle mass and the spindown expected to be within $|\dot{f}|\le 10^{-11}$\,Hz/s during most of the signals lifetime \cite{Arvanitaki:2009fg, qcdaxion, superradiance, Zhu:2020tht,LIGOScientific:2021jlr}.

Our search uses data of the LIGO H1 and L1 interferometers, from the publicly released O3a set \cite{O3aDataSet}. We limit our search to the LIGO detectors because no other detector comes within a factor of 2 of their sensitivity. 

We have reserved the O3b data \cite{O3bDataSet} for validation of any potential outliers that may persist following our ongoing follow-up investigations. This approach is commonly used for searches conducted across an extensive parameter space. Given the substantial number of trials involved, noise alone can yield outliers, and these can be reliably discarded through
verification using a separate dataset, as illustrated in \cite{Papa:2020vfz, EatHO3a}. It is important to emphasize that this procedure is applicable because we assume the presence of the same physical signal in both datasets, a
fundamental characteristic of continuous wave signals.

\section{Gravitational wave polarization data}
A rotating source such as an equatorially deformed neutron star will emit 
circularly polarized waves along the rotation axis and linearly polarized waves  perpendicular to the rotation axis. 

Gravitational wave detectors presently in operation are not omnidirectional and at any point in time there are locations on the sky where a detector is completely insensitive to some linearly polarized signals.
Because continuous wave searches aggregate data over months, the motion of the detector around the Earth and around the Sun will average the detector response and allow to detect an arbitrary source.

\begin{figure}[htbp]
\includegraphics[width=3.3in]{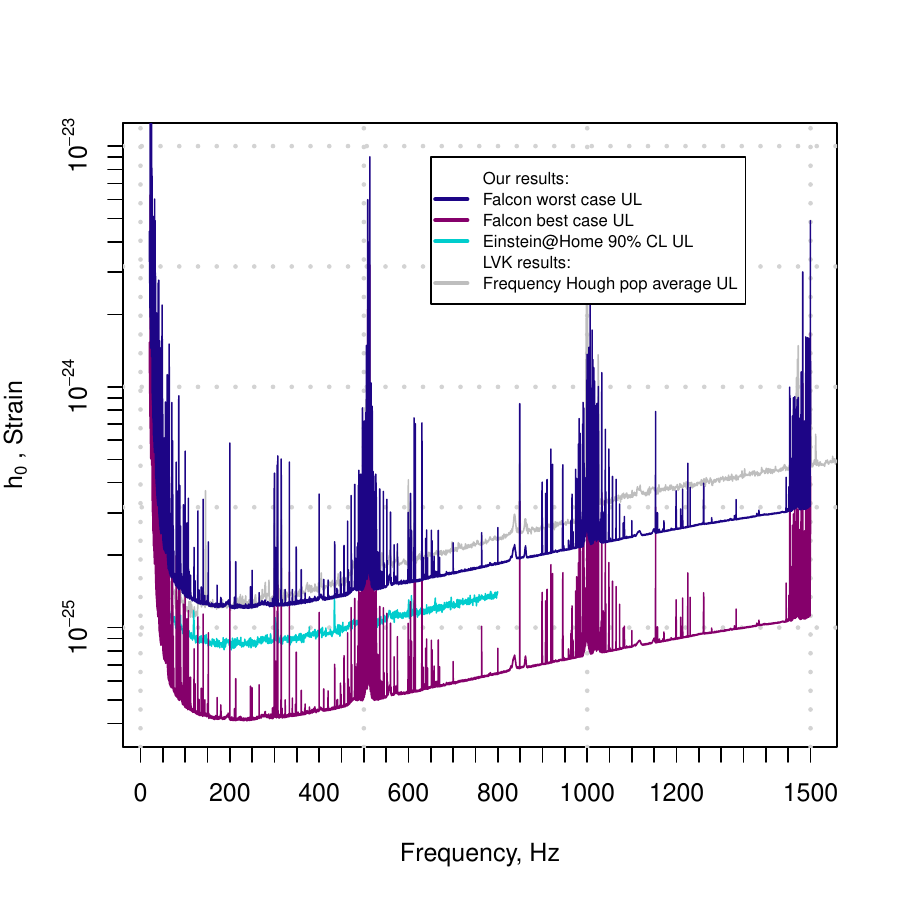}
\caption[Upper limits]{
\label{fig:amplitudeULs}
Gravitational wave intrinsic amplitude $h_0$ upper limits at 95\% confidence as a function of signal frequency. The upper limits are a measure of the sensitivity of the search. We also plot latest LIGO/Virgo and Einstein@Home all-sky population average results \cite{lvc_O3_allsky2, EatHO3a}.
}
\end{figure}

Depending on the orientation of the source to the detector line of sight, the detector will see either a single linearly polarized wave, or a mix of two. Correspondingly the amount of power received by the detector can vary by a factor of two. 
For this reason the sensitivity to continuous gravitational waves varies a lot depending on polarization as can be seen in figure \ref{fig:amplitudeULs}, with circularly polarized upper limits more than a factor of two lower than linearly polarized ones.

In the past we only reported the largest upper limit of all polarizations (worst case)  and the circular polarization upper limit. This is far from optimal if one is interested in an elliptically polarized signal - the circularly polarized limit is too small, and the worst case upper limit is unnecessarily large.

In this atlas release, we were able to include {\it functional upper limit} data \cite{functional_upper_limits}. That data consists of 14 numbers that are provided for each atlas frequency bin and each sky location. The 14 numbers are input parameters to a function of $\iota$ and $\psi$, and for every pair $(\iota, \psi)$ the function establishes a 95\% confidence upper limit. 

\section{Results}
\label{sec:results}

\begin{figure}[htbp]
\includegraphics[width=3.3in]{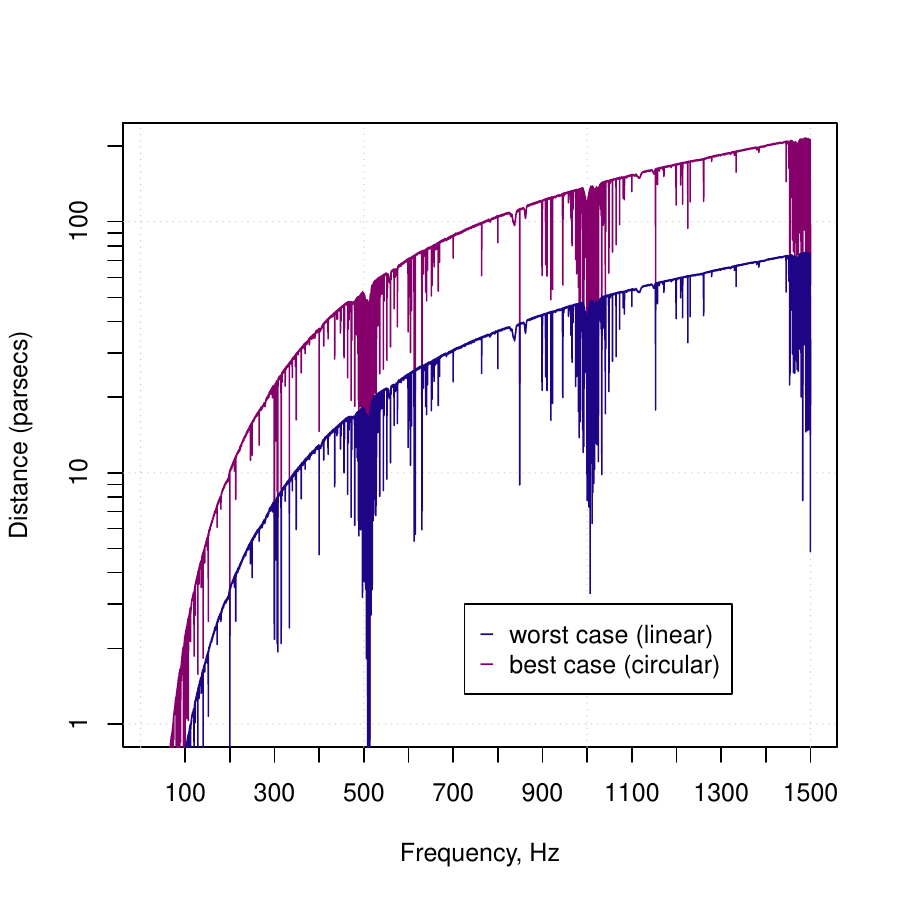}
\caption[Spindown range]{
\label{fig:distance}
Reach of the search for stars with ellipticity of $10^{-8}$. The search is also sensitive to sources with ellipticities of $10^{-7}$ with a distance from Earth that is 10 times higher. The X axis is the gravitational wave frequency, which is twice the pulsar rotation frequency for emission due to an equatorial ellipticity. R-modes and other emission mechanisms give rise to emission at different frequencies \cite{bo_rmodes}. The top curve (purple) shows the reach for a population of circularly polarized signals; The bottom curve (blue) holds for linearly polarized signals. }
\end{figure}

The Falcon pipeline calculates upper limits on the intrinsic amplitude $h_0$ of continuous gravitational waves across signal frequency and position. Our atlas includes upper limits for circularly polarized gravitational waves originating from optimally aligned sources, and worst-case upper limits derived by maximizing over polarization. These upper limits are applicable to sources in the least favorable orientation. The frequency band 20-36\,Hz is used to veto short-duration power excess stemming from disturbances. This can affect upper limits in this band, but only for exceedingly loud signals that are excluded by prior searches.

\begin{figure}[htbp]
\includegraphics[width=3.3in]{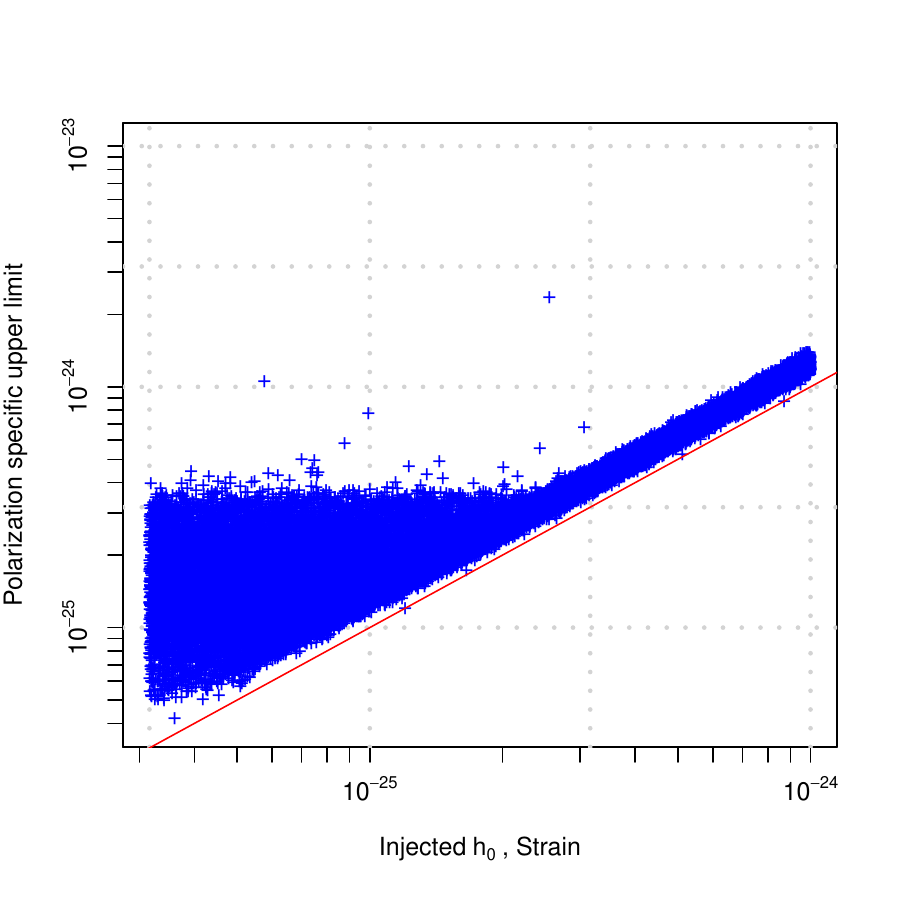}
\caption[Spindown range]{
\label{fig:uls_vs_strain}
Polarization-specific upper limits versus injected strain for a set of more than  31000 software injections. The upper limits are computed using $\iota$ and $\psi$ of the injected signals. The red line shows injected strain values.}
\end{figure}

Newly included are the functional upper limits. As this is the first implementation, there are imperfections that result in overly conservative figures. As discussed in \cite{functional_upper_limits}, the functional upper limits can be as much as 5\% larger than the original upper limits for noise-dominated data, and as much as 30\% larger for high SNR signals. Example code showing how to use functional upper limits is included in the atlas release.

The new functional upper limits have been tested with software and hardware simulated signals (``injections"). Figure \ref{fig:uls_vs_strain} shows established functional upper limits using injection-specific $\iota$ and $\psi$ versus the actual injection strength. Table \ref{tab:injections} shows the ratio of functional upper limits to injection amplitudes evaluted for $\iota$ and $\psi$ corresponding to each hardware injection \cite{hardware_injections}.

The atlas also includes the measured SNR as a function of signal frequency and source sky-position. 
Specifically, the atlas contains sky-resolved data for every $45$\,mHz signal-frequency band. The sky resolution increases proportionally to frequency, and at $1500$\,Hz at any location in the sky there is a grid point within a $0.17^\circ$ radius. 
For each sky pixel and frequency bin, we also provide the highest SNR value and the corresponding frequency, as well as $\iota$ and $\psi$ where the SNR peak has occurred.

The upper limits are summarized in Figures \ref{fig:amplitudeULs} and \ref{fig:distance}, by taking the maximum over the sky in 45 mHz frequency bands. This is the traditional way to present continuous wave upper limit results. 

The amplitude of gravitational waves received at the detector depends on the distance $d$ to the source, the source ellipticity, and the source frequency: 
\begin{equation}
h_0=\frac{4\pi^2G I_{zz} f_0^2 \varepsilon}{c^4 d},
\label{eq:epsilon}
\end{equation}
where $I_{zz}=10^{38}$\,kg\,m$^2$ is the moment of inertia of the star with respect to the principal axis aligned with the rotation axis. 

Equation \ref{eq:epsilon} suggests a convenient figure of merit to judge the sensitivity of the search, a sort of horizon distance; figure \ref{fig:distance} shows the distance to which our upper limits exclude sources with ellipticity of $10^{-8}$ as a function of frequency, for different assumed polarizations. 

This search is sensitive to sources with ellipticity of $10^{-8}$ up to 200\,pc away - an improvement on our previous results \cite{o3a_atlas1} due to expanded reach at high frequencies. 

The table \ref{tab:injections} shows the atlas results associated to  hardware injected signals, obtained by script {\tt spatial\_index\_example2.R}. The upper limits are polarization-specific and were computed using the $\iota$ and $\psi$ values of each injection. 

The frequency corresponding to the SNR peak reflects the injection frequency only when the injection is loud enough and when its parameters are within the searched area. Most injections within the searched parameter space are recovered within $0.1$\,mHz of the signal frequency. The only exception is hardware injection ip11 located in a highly contaminated low frequency band.

\begin{figure*}[htbp]
\includegraphics[width=\linewidth]{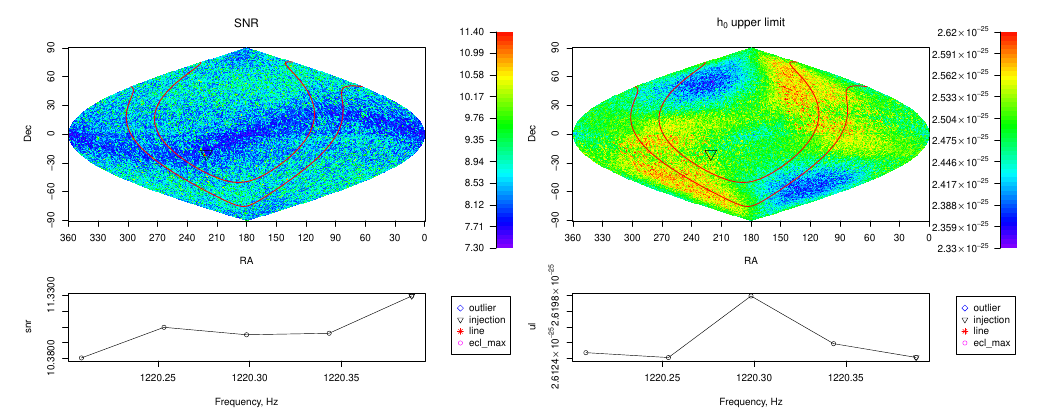}
\caption[Skymap]{
\label{fig:skymap_ul} Summary of atlas data from the bins between 1220.2-1220.4\,Hz. The top panels show the highest SNR (left) and upper limit values (right) across the frequency band, for each pixel of the sky map, using equatorial coordinates. The red lines denote the galactic plane.
The black triangle shows the location of hardware injection. The blue band of smaller SNRs near the ecliptic equator is due to large correlations between waveforms of sources in that region. The blue regions in the upper limit plot are due to the lower-SNR values in the ecliptic plane, and also occur near the ecliptic poles that are favored by the antenna pattern of the detectors. The bottom panels show the same data as a function of frequency and with the maximum taken over the sky. We mark the frequency of the band where the hardware injection was performed. The data and code used to produce this plot are available \cite{data}. 
}
\end{figure*}

Signals outside the searched parameter space are not necessarily lost. For example, despite injection ip7 being on the edge of detectability, with SNR just shy of 12 and having a large spindown, it is clearly seen in Figure \ref{fig:skymap_ul}, with the peak slightly offset from the signal location.
The established upper limit is however smaller than the injected amplitude, indicating, as expected, a loss of sensitivity  due to parameter mismatch.

The findings in \cite{lvc_O3_allsky2} span a much wider spindown range than our search. They effectively rule out neutron stars with ellipticities larger than  $10^{-6}$ within a few kiloparsecs, prompting a search for lower ellipticity sources. We accomplish this with a spindown range that covers ellipticities as high as $10^{-7}$, with higher sensitivity than \cite{lvc_O3_allsky2}. 

The atlas is made possible due to our efficient analytical upper limit procedure \cite{universal_statistics} that computes results separately for each point in the parameter space. In contrast, many searches \cite{lvc_O3_allsky2,LIGOScientific:2021jlr, keith_review, EatHO3a, Covas:2022rfg} use time-consuming Monte-Carlo simulations to  establish upper limits over the entire sky, and hence are unsuitable for the construction of an atlas.

All-sky surveys aimed at detecting continuous gravitational waves are extremely demanding in terms of computational resources. The search reported here required over 281 million CPU-hours (where CPU refers
to a single execution thread). Our atlas data provides an opportunity for others to leverage this significant investment in several ways:
\begin{itemize}
 \item 
    Followup of Astronomical Observations: The data in our atlas enables followup of astronomical observations related to pulsations. Observations from a specific celestial location can be compared
    against high Signal-to-Noise Ratio (SNR) occurrences at the same location and consistent frequencies.

 \item 
    Efficient Frequency Scanning: The atlas allows for rapid scanning across the entire frequency range at specific sky positions of interest. For example, this can be applied to the study of a supernova
    remnant, as demonstrated in the appendix, to search for significant results.

 \item 
    Ellipticity Constraints: With an estimate of the distance to a gravitational wave source, it becomes straightforward to compute upper limits on its ellipticity.

 \item 
    Predictive Modeling: Our atlas data can be directly incorporated into models of neutron star populations, facilitating predictions regarding their detectability. This same approach can be applied to
    primordial black hole populations emitting continuous gravitational waves through orbital energy loss \cite{primordial_black_holes} or tens of solar mass black hole populations generating continuous waves through super-radiance
    \cite{superradiance}.

 \item 
    Statistical Significance: The fine granularity of our atlas provides a rich dataset. In the event of coincidences with other observations, this granularity greatly aids in
    estimating the chance probability of any discovery.
\end{itemize}

\begin{table}[htbp]
\begin{center}
\begin{tabular}{l D{.}{.}{3.5} r c r r r r}
\hline
 Label & \multicolumn{1}{c}{$f$} & $\dot{f}$ & Binary & SNR & UL/$h_0$ & $\Delta f$ & In \\ 
& \multicolumn{1}{c}{Hz} & Hz/s & &  & \% & mHz & \\
\hline
\hline
ip0 & 265.57505 & -4.15e-12 & No & 28.5 & 122.5 & -0.1 & Yes\\
ip1 & 848.93498 & -3e-10 & No & 393.0 & 119.9 & -0.1 & Yes\\
ip2 & 575.16351 & -1.37e-13 & No & 39.3 & 138.5 & 0.0 & Yes\\
ip3 & 108.85716 & -1.46e-17 & No & 23.7 & 141.6 & 0.1 & Yes\\
ip4 & 1390.60583 & -2.54e-08 & No & 7.6 & 21.3 & -7.7 & No\\
ip5 & 52.80832 & -4.03e-18 & No & 155.9 & 130.2 & 0.0 & Yes\\
ip6 & 145.39178 & -6.73e-09 & No & 8.4 & 25.0 & -11.2 & No\\
ip7 & 1220.42586 & -1.12e-09 & No & 7.3 & 68.1 & 3.6 & No\\
ip8 & 190.03185 & -8.65e-09 & No & 8.9 & 83.8 & -2.9 & No\\
ip9 & 763.84732 & -1.45e-17 & No & 39.1 & 135.1 & 0.1 & Yes\\
ip10 & 26.33210 & -8.5e-11 & No & 63.9 & 124.9 & 0.0 & Yes\\
ip11 & 31.42470 & -5.07e-13 & No & 93.2 & 400.9 & -12.1 & Yes\\
ip12 & 37.75581 & -6.25e-09 & No & 14.0 & 156.5 & 4.0 & No\\
ip16 & 234.56700 & 0 & Yes & 8.3 & 29.6 & 42.7 & No\\
ip17 & 890.12300 & 0 & Yes & 8.1 & 103.6 & 23.6 & No \\
\hline
\end{tabular}

\caption[Hardware injections]{This table shows parameters of the hardware-injected  continuous wave signals and atlas data for their locations and frequencies. The upper limits for the injections are polarization specific and were computed using $\iota$ and $\psi$ of each injection. We show all the hardware injections within 20-1500\,Hz range, including those outside of our search space, as indicated by the ``In'' column. We use the reference time (GPS epoch) $t_0=1246070000$ (2019 Jul 2 02:33:02 UTC).}
\label{tab:injections}
\end{center}
\end{table}

\begin{acknowledgments}
The authors thank the scientists, engineers and technicians of LIGO, whose hard work and dedication produced the data that made this search possible.

The search was performed on the ATLAS cluster at AEI Hannover. We thank Bruce Allen, Carsten Aulbert and Henning Fehrmann for their support.

This research has made use of data or software obtained from the Gravitational Wave Open Science Center (gw-openscience.org), a service of LIGO Laboratory, the LIGO Scientific Collaboration, the Virgo Collaboration, and KAGRA. 
\end{acknowledgments}

\appendix*

\section{Using the Falcon atlas to search supernova remnants}
\label{VelaJr}
\begin{figure}[htbp]
\includegraphics[width=3.3in]{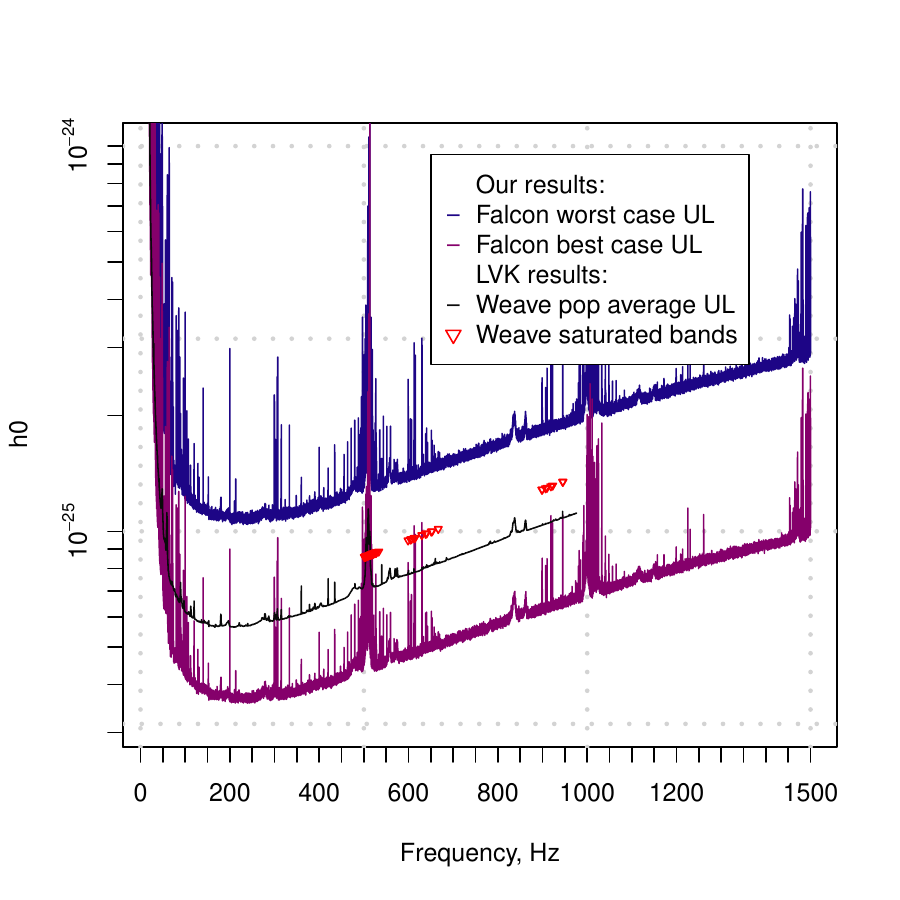}
\caption[VelaJr]{
\label{fig:velajr}
This plot shows upper limits similar to those shown in Figure \ref{fig:amplitudeULs}, but for the location of Vela Jr. Latest LIGO/Virgo/KAGRA results \cite{LVK_VelaJr} are shown for comparison. The triangles mark saturated bands for which the Weave results are invalid.
}
\end{figure}

\begin{figure}[htbp]
\includegraphics[width=3.3in]{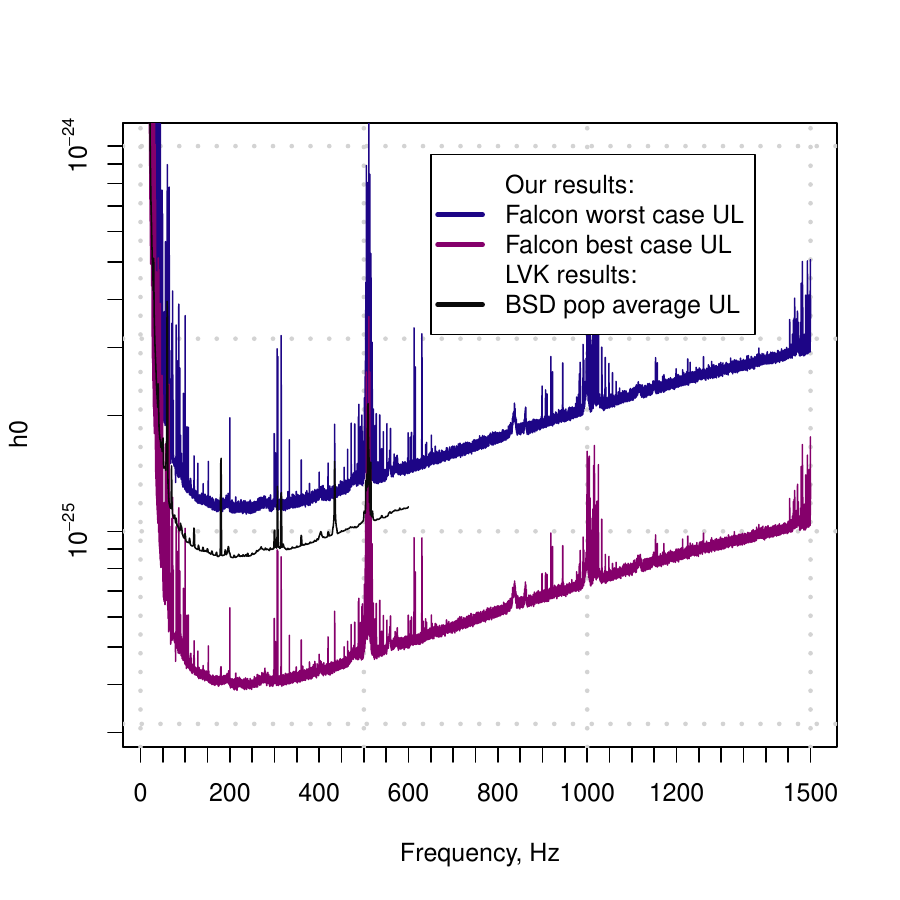}
\caption[G189]{
\label{fig:G189}
This plot shows upper limits similar to those shown in Figure \ref{fig:amplitudeULs}, but for the location of G189.1+3.0. Latest LIGO/Virgo/KAGRA results \cite{LVK_G189} are shown for comparison. The LVK upper limit curve was computed as minimum of Hanford, Livingston and Virgo data. 
}
\end{figure}

It is easy to use our atlas to examine arbitrary locations on the sky. An example R script {\tt spatial\_index\_example3.R} included with this release \cite{data} shows how to produce upper limits and SNR values for any given sky location. By simply setting right ascension and declination to the coordinates of the Vela Jr or G189.1+3.0 supernova remnants, we obtain the data shown in Figures \ref{fig:velajr} and \ref{fig:G189}, respectively. Compared to plots shown in \cite{o3a_atlas1}, our data covers a much larger frequency and spindown range.

We also plot the latest LIGO/Virgo/KAGRA results \cite{LVK_VelaJr, LVK_G189}. Our spindown range is smaller than that searched by \cite{LVK_VelaJr, LVK_G189}, but it covers observed spindowns of known pulsars. Furthermore, our results are mathematically rigorous 95\% confidence level upper limits, rather than sensitivity estimates, and we have no excluded frequency bands.

These plots illustrate that an atlas user can establish rigorous upper limits that  improve on the best dedicated directed searches.

This data can be used as is to place limits on the gravitational radiation coming from direction of these supernova remnants as we have done, or it can be the starting point for a deeper search.

\end{document}